\documentclass[onecollarge,natbib]{svjour2}
\bibpunct{[}{]}{;}{n}{}{,} 
\smartqed  
\usepackage{graphicx}
\usepackage{amsfonts}

\journalname{Few-Body Systems}
\begin{document}

\title{On a quasi-bound state in the $K^- d$ system  caused by strong interactions
\thanks{The work was supported by the Czech GACR grant 19-19640S.}
}

\titlerunning{On a quasi-bound state in the $K^- d$ system}

\author{N.V. Shevchenko 
}


\institute{N.V. Shevchenko \at
              Nuclear Physics Institute, 25068 \v{R}e\v{z}, Czech Republic \\
              \email{shevchenko@ujf.cas.cz} 
}

\date{Received: date / Accepted: date}

\maketitle

\begin{abstract}
It was found that $NN$ potential could influence the results of the quasi-bound state
search in the $K^- d$ system, where the corresponding pole is situated close to
the threshold. Three-body Faddeev-type calculations of the $\bar{K}NN - \pi \Sigma N$
system performed with a new model of nucleon-nucleon interaction predict
the existence of the quasi-bound $K^- d$ state caused by strong interactions. Its
binding energy is small ($1-2$ MeV), while the width is comparable with the width of
the $K^- pp$ quasi-bound state ($40-60$ MeV).

\keywords{Few-body exotic systems \and AGS equations \and Antikaon-nucleon interaction}
\end{abstract}


\section{Introduction}
\label{intro_sect}

Many theoretical works and several experiments were devoted to the question
of a quasi-bound state in systems consisting of an antikaon(s) and nucleons,
 see e.g. \cite{review}. Particular interest was attracted to the lightest possible system
$\bar{K}NN$ with zero spin (usually denoted as $K^- pp$): all theoretical works
predicted the quasi-bound state in it. The methods of and the inputs
for the calculations are quite different, and the predicted binding energy and width
of the state also variate in quite a wide range. We also predicted a quasi-bound
state in the spin zero $\bar{K}NN$ system \cite{review} with binding
energy and width strongly depending on the particular form of the antikaon-nucleon potential,
which is used as a input. We solved three-body Faddeev-type AGS equations \cite{AGS3}
with coupled $\bar{K}NN$ and $\pi \Sigma N$ channels and used three different models
of $\bar{K}N$ interaction: two phenomenological potentials and one chirally motivated model.

We also studied another state of the $\bar{K}NN$ system with spin one (which will be denoted
as $K^- d$). It has an atomic state, kaonic deuterium, which is mainly caused by Coulomb
attraction between $K^-$ and $p$, while the strong interactions give corrections to the binding
energy and width (see \cite{Revai_Kdeu} and references therein). 
But we did not find a pole caused by purely strong interactions corresponding to
the quasi-bound state in $K^- d$ \cite{ourKNN_I} similar to that one in the $K^- pp$ system.
We demonstrated, that the strong quasi-bound state appears
in the $K^- d$ system if the attraction in the isospin-zero $\bar{K}N-\bar{K}N$
part of the coupled-channel $\bar{K}N - \pi \Sigma$ potential is increased
by hands.

After studying $\bar{K}NN$ and $\bar{K}\bar{K}N$ systems (the last one also has a quasi-bound
state \cite{KKN}) we started investigations  of the four-body $\bar{K}NNN$ system. We are using
four-body Faddeev-type equations \cite{AGS4} on four-body transition amplitudes, which
contain, among others, three-body transition amplitudes evaluated at every step of
the four-body calculations. We constructed a new nucleon-nucleon potential
for these calcula\-tions, and found out that the quasi-bound state caused by the strong
interactions appeared in the  $K^- d$ system. The present paper is devoted to this state.

\section{Three-body AGS equations and two-body input}
\label{Method_sect}

The calculations of the quasi-bound state in the $K^- d$ system caused by strong
interactions were performed using the same method and input as those used
for the $K^- pp$ system \cite{review}. We solved Faddeev-type three-body equations in
AGS form \cite{AGS3} with coupled $\bar{K}NN$ and $\pi \Sigma N$
channels. The three-body equations in operator form for the $\bar{K}NN$ system with spin
one ($K^- d$) are those for the spin-one $\bar{K}NN$ ($K^- pp$): Eqs. (23--25) of \cite{review}.
Written in momentum representation they differ from the $K^- pp$ case by the antisymmetrization
and $6j$ recoupling coefficients.

The three-body AGS equations in momentum representation are integral equations. We solved
them in two ways. One of them is the direct search of the pole position in the complex
plane. Another one is the $1/|{\rm Det}|^2$ method proposed and successively used
for the $K^- pp$ system  in \cite{ourKNN_II} . The idea of the method is that a pole,
situated under the threshold, must be seen at the real energy axis as a bump. The position
and the width of the quasi-bound state can be evaluated by fitting the bump by Breit-Wigner formula
with some arbitrary background. The two methods supplement each other: the direct search needs
initial values for the eigenvalue problem solving, and they can be provided by
the $1/|{\rm Det}|^2$ method. On the other hand, the binding energy and width obtained from bump
fitting could be used as a control of the direct search result.

The input for the AGS equations describing the $\bar{K}NN - \pi \Sigma N$ system are $T$-matrices
corresponding to the $\bar{K}N - \pi \Sigma$, $NN$, $\Sigma N$ and $\pi N$ potentials. 
All the potentials have separable form, which simplifies the three-body equations,
all of them reproduce low-energy experimental data for the corresponding subsystem
quite accurately. Keeping in mind uncertainties in experimental data used for
construction of the interaction models, we see no reason to introduce 
some three-body force and by  this increase ambiguity of the calculations.
Coulomb interaction between $K^-$ and $p$ was not included into the equations,
since in contrast to atomic state calculations, where Coulomb plays the main role,
the present work is devoted to the quasi-bound $K^- d$ state caused by strong interactions,
where Coulomb should lead to some small corrections only.

Interaction of an antikaon with a nucleon plays the main role in the calculations.
It is strongly attractive in the isospin zero state, which had lead to the question
of a quasi-bound state in few- and/or many-body systems consisting of antikaon(s)
and nucleons existence. The $\bar{K}N$ system is coupled to the $\pi \Sigma$ channel through
the $\Lambda(1405)$ resonance, which is usually assumed as a resonance in the
$\pi \Sigma$ channel and a quasi-bound state in the $\bar{K}N$ channel.
The resonance is formed by one or two poles (this question rose quite vivid discussions)
and, according to the Particle Data Group \cite{PDG} has a mass $1405.1^{+ 1.3}_{- 1.0}$
MeV and a width $50.5 \pm 2.0$ MeV.  

We used three our models of antikaon-nucleon interaction, constructed and used before.
They are: two phenomenological potentials with coupled $\bar{K}N$ and $\pi \Sigma$
channels and one- or two-pole structure of the $\Lambda(1405)$ resonance
($V_{\bar{K}N}^{\rm 1,SIDD}$ and $V_{\bar{K}N}^{\rm 2,SIDD}$ correspondingly),
and the chirally motivated $\bar{K}N - \pi \Sigma - \pi \Lambda$ potential
$V_{\bar{K}N}^{\rm Chiral}$ (with two-pole $\Lambda(1405)$ structure).
Parameters of the phenomenological and the chirally motivated  potentials 
are presented in \cite{VKN_SIDD12} and  \cite{ourKNN_I} correspondingly. 
Yamaguchi form-factors were used for the one-pole phenomenological and chirally
motivated potentials together with the $\bar{K}N$ form-factor of the two-pole
phenomenological potential. The $\pi \Sigma$ form-factor of the two-pole
$V_{\bar{K}N}^{\rm 2,SIDD}$ has more complicated form. The chirally-motivated
potential is energy-dependent one.

All three potentials were fitted to the experimental data on kaonic hydrogen and
low-energy $K^- p$ scattering. In particular, they reproduce $1s$ level shift and width
in kaonic hydrogen, caused by the strong interaction additional to the main Coulomb
interaction, measured by SIDDHARTA experiment \cite{SIDDHARTA}. In contrast
to other authors we reproduce these observables directly without using a Deser-like
approximate formula connecting the characteristics of kaon hydrogen with the $K^- p$
scattering length. Our three models of the antikaon-nucleon interaction also reproduce
elastic and inelastic cross-sections of $K^- p$ scattering
\cite{Kp2exp,Kp3exp_1,Kp3exp_2,Kp4exp,Kp5exp,Kp6exp} together with threshold branching
ratios $\gamma$, $R_c$ and $R_n$ \cite{gammaKp1,gammaKp2} (or $R_{\pi \Sigma}$ 
constructed from the last two in the case of phenomenological potentials).

The phenomenological models with coupled $\bar{K}N$ and $\pi \Sigma$ channels
were used in the three-body coupled-channel $\bar{K}NN - \pi \Sigma N$ calculations,
while the corresponding exact optical $\bar{K}N(-\pi \Sigma)$  potentials\footnote{The exact
optical potential is the one-channel potential, which provides exactly the same elastic amplitude
as the coupled-channel model of interaction (see e.g. \cite{my_Kd}).}
were taken in approximate one-channel $\bar{KNN}$ calculations. The chirally motivated
$\bar{K}N -\pi \Sigma-\pi \Lambda$ interaction model was used as
$\bar{K}N - \pi \Sigma (-\pi \Lambda)$ in the three-body calculation with coupled channels
since three-body channel with $\Lambda$ is not included into the equations. The exact optical
$\bar{K}N (-\pi \Sigma - \pi \Lambda)$ version of interaction entered the approximate three-body
equations.

Spin dependent and spin independent potentials of the $\Sigma N$ interaction
were constructed in \cite{my_Kd}. They are one-term separable potentials with Yamaguchi
form-factors.The parameters were fitted to the experimental data on $\Sigma N$ and
$\Lambda N$  cross-sections \cite{SigmaN1,SigmaN2,SigmaN3,SigmaN4,SigmaN5}.
Isospin $1/2$ $\Sigma N - \Lambda N$ potential is
a two-channel one, it is used in the three-body calculations in a form of the exact
optical $\Sigma N (-\Lambda N)$ potential. The isospin $3/2$ $\Sigma N$ potential
has only one channel. In the present calculations we used the $\Sigma N$ interaction
model, which does not depend on spin.

The $\pi N$ potential is assumed to play negligible role, and was omitted. Finally, the new
nucleon-nucleon potential, which lead to the $K^- d$ quasi-bound state appearance, is
described in the next section.

\section{New separable $NN$ potential}
\label{V_NN_sect}

The Two-term Separable New potential (TSN) of nucleon-nucleon interaction has a form
\begin{equation}
\label{VNN}
V_{NN}^{\rm TSN}(k,k') = \sum_{m=1}^2 g_{m}(k) \, \lambda_{m} \; g_{m}(k') \,,
\end{equation}
with form-factors
\begin{equation}
\label{gNN}
g_{m}(k) = \sum_{n=1}^3 \frac{\gamma_{mn}}{(\beta_{mn})^2 + k^2},
\quad {\rm for \;} m=1,2.
\end{equation}
It was fitted to Argonne $V18$ potential \cite{ArgonneV18} phase shifts without the condition,
which was imposed on the previously used $V_{NN}^{\rm TSA}$ potential \cite{Doles_NN}:
one of the normalization constants of the deuteron way function must be equal to zero.
Technically the new $V_{NN}^{\rm TSN}$ and previously used $V_{NN}^{\rm TSA}$ models
differ by the number of terms in the form-factors, Eq.(\ref{gNN}).  The parameters of
the new two-term separable potential are presented in Table \ref{VNN_np.tab} 
for the triplet and in Table \ref{VNN_pp.tab} for the singlet case.
\begin{table}[hb]
\caption{Parameters of the new $V_{NN}^{\rm TSN}$ potential, triplet:
strength constants $\lambda_m$, range $\beta_{mn}$ and additional
$\gamma_{mn}$ parameters.}
\label{VNN_np.tab}
\begin{center}
\begin{tabular}{cccccccc}
\hline \hline \noalign{\smallskip}
{} &\, $\lambda_{m}$ &\, $\beta_{m1}$  &\, $\beta_{m2}$  &\, $\beta_{m3}$
&\, $\gamma_{m1}$  &\, $\gamma_{m2}$  &\, $\gamma_{m3}$ \\
\noalign{\smallskip} \hline \noalign{\smallskip}
$m=1$ &\, $-1.9938$ &\, $1.2096$ &\, $3.2135$ &\, $1.3912$
 &\quad $0.0884$ &\quad $1.9889$ &\, $-0.1027$  \\
$m=2$ &\quad$  1.7584$ &\, $3.9940$ &\, $3.9999$ &\, $2.7070$
 &\, $-1.9660$ &\, $-1.9225$ &\quad $0.4144$  \\
\noalign{\smallskip} \hline \hline
\end{tabular}
\end{center}
\end{table}

Four parameters $(\lambda_m, \gamma_{m1},\gamma_{m2},\gamma_{m3})$  for every $m$
can be replaced by three independent parameters
$(\lambda'_m=\lambda_m*\gamma_{m3}^2$, $\gamma'_{m1}=\gamma_{m1}/\gamma_{m3}$,
$\gamma'_{m2}=\gamma_{m2}/\gamma_{m3})$, correspondingly, with $\gamma'_{m3}=1$.
Therefore, $np$ and $pp$ potentials contain $12$ independent parameters each.
Such transformation does not change the potential Eq.(\ref{VNN}) and can be done with
$\gamma_{m1}$ or $\gamma_{m2}$ as well. However, the ''natural'' values
of the parameters, presented in Tables \ref{VNN_np.tab} and \ref{VNN_pp.tab}
are more convenient for numerical calculations. 

The triplet and singlet scattering lengths $a$ and effective ranges $r_{\rm eff}$
provided by the TSN potential are:
\begin{eqnarray}
a_{np}^{\rm TSN} = -5.400  {\, \rm fm}, \qquad  r_{{\rm eff}, np}^{\rm TSN} = 1.744 {\, \rm fm} \\
a_{pp}^{\rm TSN} = 16.325 {\, \rm fm}, \qquad  r_{{\rm eff}, pp}^{\rm TSN} = 2.792 {\, \rm fm,}
\end{eqnarray}
the binding energy of the deuteron is $E_{\rm{deu}} = 2.2246$ MeV. The previously
used TSA potential (its TSA-B version) provides slightly different scattering lengths
and effective ranges:
\begin{eqnarray}
a_{np}^{\rm TSA-B} = -5.413  {\, \rm fm}, \qquad  r_{{\rm eff}, np}^{\rm TSA-B} = 1.760 {\, \rm fm} \\
a_{pp}^{\rm TSA-B} = 16.559 {\, \rm fm}, \qquad  r_{{\rm eff}, np}^{\rm TSA-B} = 2.880 {\, \rm fm}
\end{eqnarray}
and the same binding energy of deuteron ($2.2246$ MeV).
\begin{table}[hb]
\caption{Parameters of the new $V_{NN}^{\rm TSN}$ potential, singlet:
strength constants $\lambda_m$, range $\beta_mn$ and additional
$\gamma_{mn}$ parameters.}
\label{VNN_pp.tab}
\begin{center}
\begin{tabular}{cccccccc}
\hline \hline \noalign{\smallskip}
{} &\, $\lambda_{m}$ &\, $\beta_{m1}$  &\, $\beta_{m2}$  &\, $\beta_{m3}$
&\, $\gamma_{m1}$  &\, $\gamma_{m2}$  &\, $\gamma_{m3}$ \\
\noalign{\smallskip} \hline \noalign{\smallskip}
$m=1$ &\, $ -1.9793$ &\, $1.8855$ &\, $2.8396$ &\, $1.1834$
 &\, $  -0.1800$ &\quad $1.9999$ &\, $0.0362$  \\
$m=2$ &\quad $1.7815$ &\, $3.9897$ &\, $3.9919$ &\, $0.5000 $
 &\, $-1.8881$ &\, $-1.9914$ &\, $0.0014$  \\
\noalign{\smallskip} \hline \hline
\end{tabular}
\end{center}
\end{table}

\begin{figure}
\centering
\includegraphics[width=0.9\textwidth]{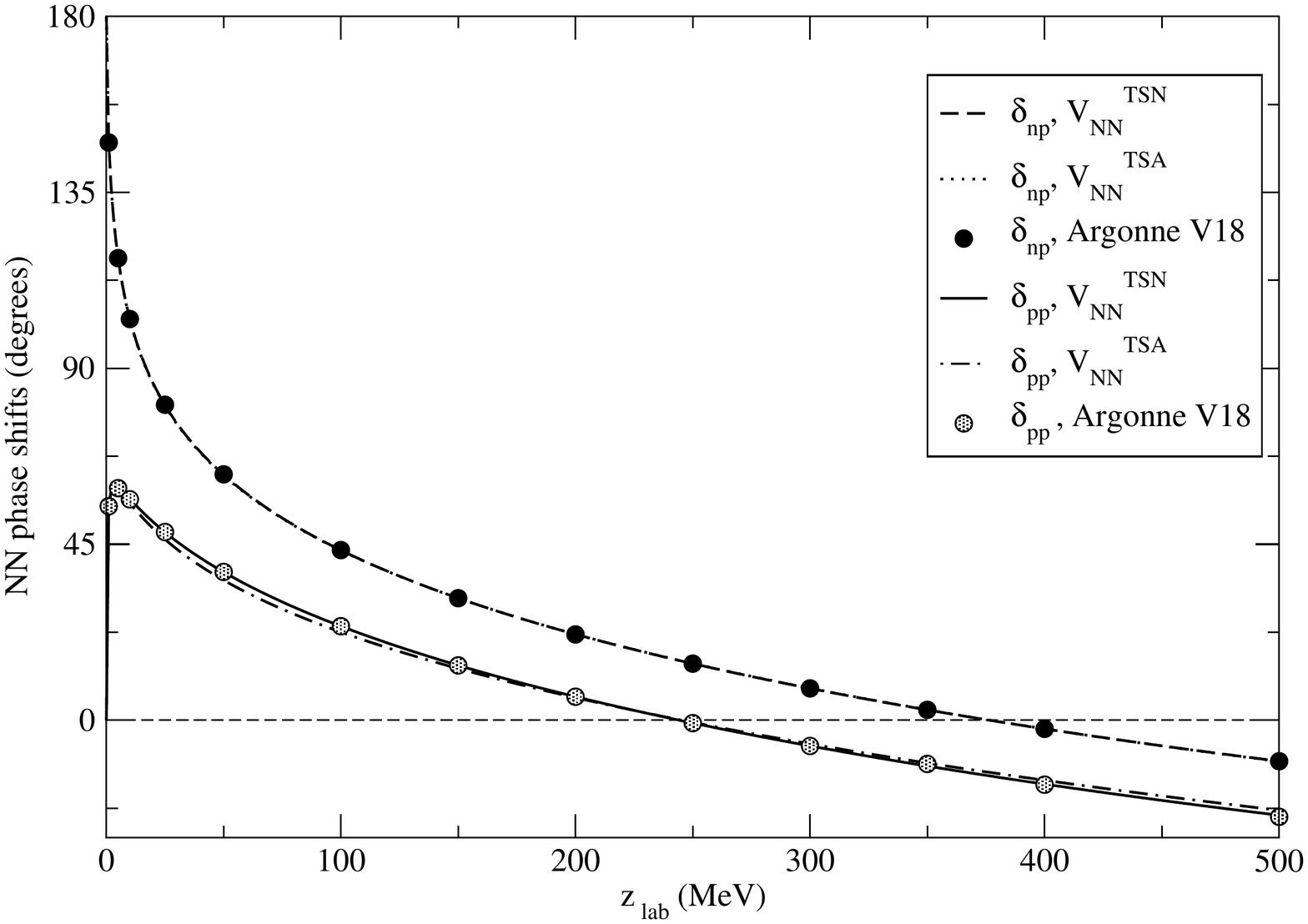}
\caption{Phase shifts of $np$ and $pp$ scattering calculated using the new $V^{\rm TSN}_{NN}$
and previously used $V^{\rm TSA-B}_{NN}$ potentials (lines) compared with phase
shifts of Argonne $V18$ potential (circles).}
\label{NNphases.fig}
\end{figure}

Phase shifts of $np$ and $pp$ scattering calculated using the new $V_{NN}^{\rm TSN}$ potential
are plotted together with those calculated using TSA-B version of the previously used
model in Figure \ref{NNphases.fig}. It is seen that the phase shifts given by Argonne $V18$
potential \cite{ArgonneV18} (black circles for $\delta_{np}$ and circles filled with dots for
$\delta_{pp}$) are reproduced by the new potential with high accuracy. The phases of the new
TSN and the previously used TSA-B potentials are practically indistinguishable for the case of
$\delta_{np}$ (dashed and dotted lines), while $pp$ phases are slightly different (solid and
dash-dotted lines for TSN and TSA-B potentials correspondingly). The phases change their signs, which
means that all three potentials are repulsive at short distances.

\section{Results and discussion}
\label{results_sect}

The results of the direct search of the pole in the $K^- d$ system  corresponding
to the quasi-bound state state caused by strong interactions are shown in Table \ref{polesKd.tab}.
The new nucleon-nucleon $V_{NN}^{\rm TSN}$ potential was used together with
the three models of antikaon-nucleon interaction. The calculations with coupled
$\bar{K}NN - \pi \Sigma N$ channels using the phenomenological $\bar{K}N$ 
potential with one-pole structure of the $\Lambda(1405)$ resonance
$V_{\bar{K}N}^{\rm 1,SIDD}$ provide no quasi-bound state.
In contrast to it, the models with two-pole structure: the phenomenological
$V_{\bar{K}N}^{\rm 2,SIDD}$ and chirally-motivated $V_{\bar{K}N}^{\rm Chiral}$
potentials, - provide a quasi-bound state with small binding energy $B_{K^- d}$.
The energy is counted from the $K^- d$ threshold
$z_{th, {\rm K^- d}} = m_{\bar{K}} + 2 m_N + E_{\rm deu}$ $=2371.26$ MeV.
\begin{table}[h]
\caption{Binding energy $B_{K^-d}$ (MeV) and width $\Gamma_{K^- d}$ (MeV)
of the quasi-bound state in the $K^- d$ system calculated using direct pole
search in the complex energy plane. The results obtained by coupled-channel 
$\bar{K}NN - \pi \Sigma N$ equations solving and one-channel $\bar{K}NN$ variant
with exact optical antikaon-nucleon potentials are presented. Phenomenological $\bar{K}N$
potentials with one-pole $V_{\bar{K}N}^{\rm 1,SIDD}$ and two-pole $\Lambda(1405)$ structure
$V_{\bar{K}N}^{\rm 2,SIDD}$ were used together
with the chirally-motivated model $V_{\bar{K}N}^{\rm Chiral}$ of the antikaon-nucleon interaction.
The binding energy is counted from the threshold energy of the
$K^- d$ system:  $z_{th, {\rm K^- d}} = m_{\bar{K}}+2 \, m_N + E_{\rm deu}$ $=2371.26$ MeV.}
\label{polesKd.tab}
\begin{center}
\begin{tabular}{ccccc}
\hline \hline \noalign{\smallskip}
 & \multicolumn{2}{c}{Coupled-channels} 
    & \multicolumn{2}{c}{With exact optical} \\
 & \multicolumn{2}{c}{calculation} 
    & \multicolumn{2}{c}{$\bar{K}N$ potential} \\
\hline \noalign{\smallskip}
  & $B_{K^- d}$  & $\Gamma_{K^-d}$  & $B_{K^- d}$  & $\Gamma_{K^-d}$ \\
\noalign{\smallskip} \hline \noalign{\smallskip}
 $V_{\bar{K}N}^{\rm 1,SIDD}$ & $-$ & $-$ & $0.8$ & $68.3$ \\
\noalign{\smallskip} 
 $V_{\bar{K}N}^{\rm 2,SIDD}$ & $0.9$ & $59.4$ & $3.8$ & $63.2$ \\
\noalign{\smallskip} 
 $V_{\bar{K}N}^{\rm Chiral}$   & $1.3$ & $41.8$ & $0.9$ & $43.6$ \\
\noalign{\smallskip} \hline \hline
\end{tabular}
\end{center}
\end{table}

No $K^- d$ quasi-bound state was found in our previous work \cite{ourKNN_I}, where similar
calculations were performed with $V_{NN}^{\rm TSA}$ potential. Therefore the particular model
of nucleon-nucleon interaction, not only of antikaon-nucleon one, can influence the result
of the quasi-bound pole search in the $K^- d$ system.

It was demonstrated in \cite{ourKNN_I} that gradual increasing of the absolute value
of the isospin zero constant $\lambda_{I=0}^{\bar{K}\bar{K}}$  of the  $\bar{K}N - \pi \Sigma$
potential leads to the quasi-bound state appearance in the $K^- d$ system. And the necessary
changes are small, as minimum, for the system described by the two-pole phenomenological
$\bar{K}N$ potential. Indeed, it is seen at Fig. 6 of \cite{ourKNN_I}, that the pole
calculated with $V_{\bar{K}N}^{\rm 2,SIDD}$ potential with multiplication factor $1$ is situated
slightly above the $K^- d$ threshold, so that the $K^- d$ system is almost bound. On the other
hand, we have shown in \cite{my_Kd}, that $NN$ interaction plays a minor role in $K^- d$
scattering length calculations. It is seen at Fig. 12 of \cite{my_Kd} that use of PEST nucleon-nucleon
potential which is not repulsive at short distances changes $a_{K^- d}$ only slightly in comparison
with the results obtained with $V_{NN}^{\rm TSA}$ (with is repulsive at short distances).
Keeping all this in mind, the few MeV binding energy of the $K^-  d$ quasi-bound state
caused by strong interactions is an expected result. The positions of the poles calculated
with the previously used nucleon-nucleon $V_{NN}^{\rm TSA}$ potential
are so close to the $K^- d$  threshold (from above), that use of another $V_{NN}$ pushes
the poles downward under the threshold, so that the quasi-bound state appeared.

The widths of the $K^- d$ quasi-bound states presented in Table  \ref{polesKd.tab} confirm
our suggestions \cite{ourKNN_I} that they must be comparable with those for the $K^- pp$
quasi-bound state.

The results of the approximate calculations performed with the exact optical versions of the three
antikaon-nucleon potentials are also shown in Table \ref{polesKd.tab}. In that case 
the quasi-bound state also appeared in the $K^- d$ system described by the one-pole
$V_{\bar{K}N}^{\rm 1,SIDD}$ potential in addition to the two others. The binding energies of
the already existing $K^- d$ quasi-bound states (described by the two-pole $V_{\bar{K}N}$)
has changed, and they become broader.

An atomic state, kaonic deuterium (see \cite{Revai_Kdeu} and references therein)
exists in the spin one $\bar{K}NN$ system, however it cannot be misidentified
with the quasi-bound state caused by the strong interactions. Indeed, the binding energy
of kaonic deuterium is $\sim 10$ keV, while strong interactions bound the system
with $1-2$ MeV. The differences in widths of the atomic and strong $K^- d$ states
is even more drastic: tens of MeV for the strong quasi-bound states in contrast to $\sim 1$ keV
for kaonic deuterium \cite{Revai_Kdeu}.

\begin{figure}[b]
\centering
\includegraphics[width=0.9\textwidth]{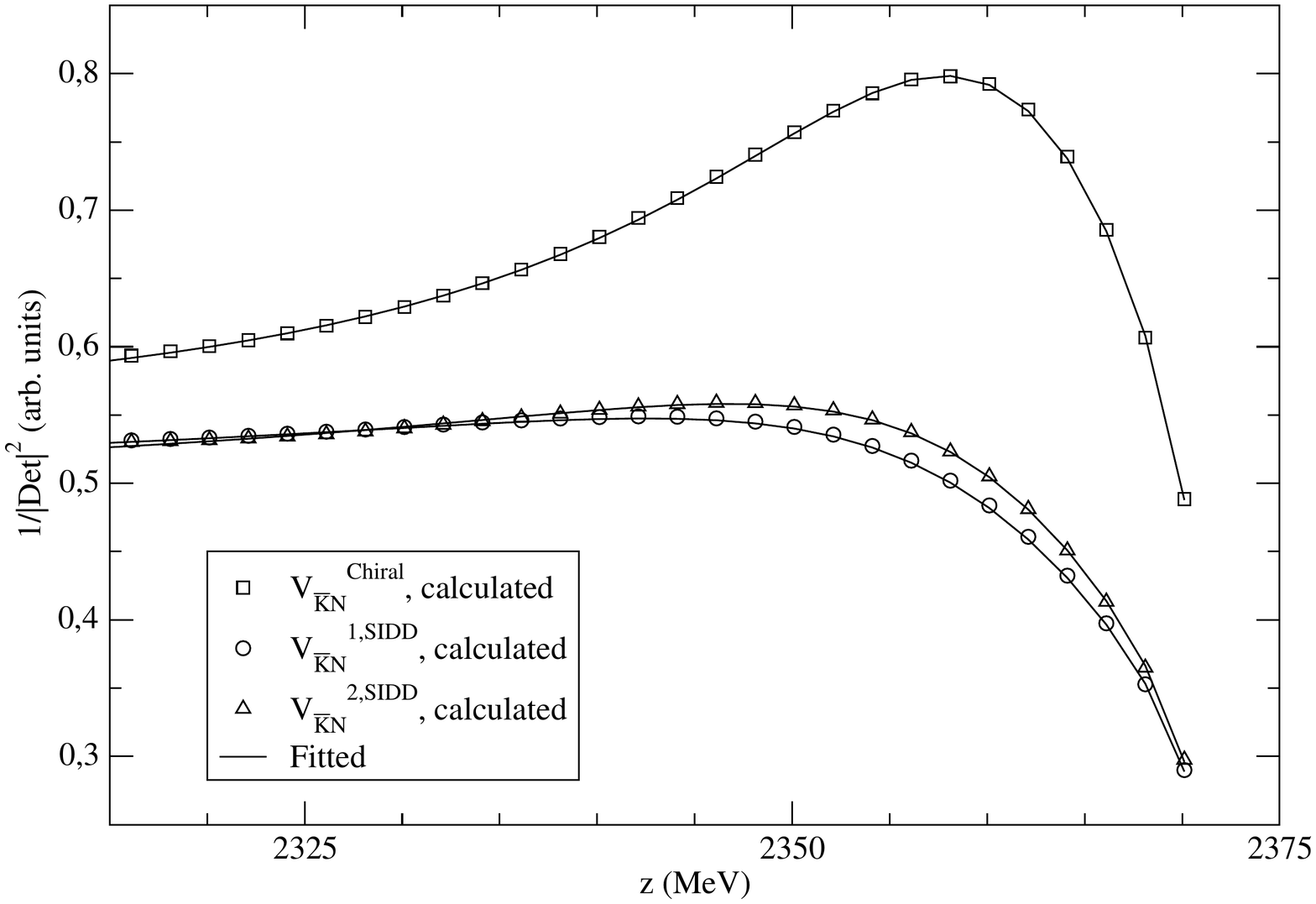}
\caption{Evaluated $1/|{\rm Det(z)}|^2$ functions for the $K^- d$ system
calculated using chirally-motivated $V_{\bar{K}N}^{\rm Chiral}$ (squares),
one-pole phenomenological $V_{\bar{K}N}^{\rm 1,SIDD}$ (circles) and two-pole
phenomenological $V_{\bar{K}N}^{\rm 2,SIDD}$ (triangles) antikaon-nucleon
potentials. The lines are the corresponding non-linear fits of the functions.}
\label{DetfitsKd.fig}
\end{figure}
We also performed calculations of the $1/|\rm Det|^2$ functions, which provided quite
good results for the $K^- pp$ system \cite{ourKNN_II}. The bumps corresponding
to the $K^- pp$ poles evaluated with the three models of antikaon-nucleon interaction
were clearly pronounced and have a resonance-like form. On the contrary, the $K^- d$
bumps in the $1/|{\rm Det}|^2$ functions are situated very close to the $\bar{K}NN$ threshold
and are not clearly pronounced, especially those obtained with the phenomenological potentials.
It is seen in Figure \ref{DetfitsKd.fig}, where the points calculated using one-pole
phenomenological $V_{\bar{K}N}^{\rm 1,SIDD}$ (circles), two-pole phenomenological
$V_{\bar{K}N}^{\rm 2,SIDD}$ (triangles) and chirally motivated $V_{\bar{K}N}^{\rm Chiral}$
(squares) potentials are joined by the fitting lines. 
Breit-Wigner fits of the calculated $1/|{\rm Det}|^2$ functions with arbitrary background
in the $K^- d$ system are far not so accurate as in the case of $K^- pp$ system:
\begin{equation}
\begin{tabular}{ll}
$B_{K^- d, BW}^{\rm 1,SIDD} =  9.2$ \, MeV,  \quad  & 
  $\Gamma_{K^-d, BW}^{\rm 1,SIDD} = 59.6$ \, {\rm MeV,}  \\[1mm]
$B_{K^- d, BW}^{\rm 2,SIDD} = 11.4$ \, MeV, \quad &
  $\Gamma_{K^-d, BW}^{\rm 2,SIDD} = 52.2$ \, {\rm MeV,} \\[1mm]
$B_{K^- d, BW}^{\rm Chiral} =  5.3$ \, MeV,   \quad & 
  $\Gamma_{K^-d, BW}^{\rm Chiral} =  48.6$ \, {\rm MeV.} 
\end{tabular}
\end{equation}
It seems that the $1/|{\rm Det}|^2$ method is not good for quasi-bound states
situated so close to the threshold.

For comparison we studied, how the new NN potential changes binding energies and widths
of the $K^- pp$ quasi-bound states. The results are presented in Table \ref{polesKpp.tab}, where
the characteristics of the $K^- pp$ quasi-bound state calculated using the system of equations
with coupled $\bar{K}NN$ and $\pi \Sigma N$ channels are shown together with the results
of one-channel calculations with exact optical $\bar{K}N$ potentials. The $K^- pp$ 
binding energies and widths obtained with the previously use TSA-B NN potential
and published in~\cite{ourKNN_II} are also shown. The binding energy is counted from
the threshold energy of the $K^-pp$ system:  $z_{th, {\rm K^- pp}} = m_{\bar{K}}+2 \, m_N$ 
$=2373.485$ MeV. It is seen that the new model of
NN interaction also changed the pole positions on few MeV. However, since the quasi-bound
state in the $K^- pp$ system, in contrast to $K^- d$, is situated far from the $\bar{K}NN$
and $\pi \Sigma N$ thresholds, these changes do not have so drastic effect.
\begin{table}[ht]
\caption{Binding energy $B_{K^-pp}$ (MeV) and width $\Gamma_{K^- pp}$ (MeV)
of the quasi-bound state in the $K^- pp$ system calculated using direct pole
search in the complex energy plane. The results obtained by coupled-channel 
$\bar{K}NN - \pi \Sigma N$ equations solving and one-channel $\bar{K}NN$ variant
with exact optical antikaon-nucleon potentials are presented. Phenomenological $\bar{K}N$
potentials with one-pole $V_{\bar{K}N}^{\rm 1,SIDD}$ and two-pole $\Lambda(1405)$ structure
$V_{\bar{K}N}^{\rm 2,SIDD}$ were used together
with the chirally-motivated model $V_{\bar{K}N}^{\rm Chiral}$ of the antikaon-nucleon interaction.
Previous results from \cite{ourKNN_II} are also shown.
The binding energy is counted from the threshold energy of the
$K^-pp$ system:  $z_{th, {\rm K^- pp}} = m_{\bar{K}}+2 \, m_N$ $=2373.485$ MeV.}
\label{polesKpp.tab}
\begin{center}
\begin{tabular}{ccccccc}
\hline \hline \noalign{\smallskip}
  & \multicolumn{2}{c}{Coupled channels} 
    & \multicolumn{2}{c}{With exact optical} 
     & \multicolumn{2}{c}{Previous results} \\
  & \multicolumn{2}{c}{calculation}  
   & \multicolumn{2}{c}{$\bar{K}N$ potential} 
    & \multicolumn{2}{c}{from \cite{ourKNN_II}} \\
\hline \noalign{\smallskip}
  & $B_{K^- pp}$  & $\Gamma_{K^- pp}$  & $B_{K^- pp}$  & $\Gamma_{K^-pp}$   
    & $B_{K^- pp}$  & $\Gamma_{K^-pp}$ \\
\noalign{\smallskip} \hline \noalign{\smallskip}
 $V_{\bar{K}N}^{\rm 1,SIDD}$ & $52.18$ & $67.1$ & $53.29$ & $63.3$ & $53.29$ & $64.9$ \\
\noalign{\smallskip} 
$V_{\bar{K}N}^{\rm 2,SIDD}$ & $46.56$ & $51.2$ & $46.65$ & $47.4$ & $47.45$ & $49.8$ \\
\noalign{\smallskip} 
 $V_{\bar{K}N}^{\rm Chiral}$    & $29.43$ & $46.4$ & $30.01$ & $46.6$ & $32.24$ & $48.6$\\
\noalign{\smallskip} \hline \hline
\end{tabular}
\end{center}
\end{table}

Comparing the $K^- d$ and $K^- pp$ characteristics of the quasi-bound states caused
purely by strong interactions  in Tables \ref{polesKd.tab} and \ref{polesKpp.tab} correspondingly,
we see large difference between the binding energies in the both systems: the $K^- pp$
is bound much stronger than $K^- d$. The $K^- d$ and $K^- pp$ widths are comparable. We see,
that the particular model of $V_{NN}$ plays a minor role in the $K^- pp$ system, differences in
the pole positions (below and above the threshold) in the $K^- d$ system are also not very big.
However, since the $K^- d$ poles are situated very close to the $K^- d$ threshold, these small
differences resolve the question of the quasi-bound state existence.

\section{Summary}
\label{Summary_sect}

We found that $NN$ potential could influence the results of the quasi-bound state
search in the $K^- d$ system, where the corresponding pole is situated close to the $K^- d$ threshold,
and predicted possibility of the $K^- d$ quasi-bound state existence caused by the strong interactions.
Three-body Faddeev-type calculations performed with the new TSN model  of nucleon-nucleon 
interaction found out the quasi-bound state in this system with binding energy $1-2$ MeV and
width comparable with those obtained for the $K^- pp$ system, $40-60$ MeV. The quasi-bound
state caused by strong interactions is stronger bound and is much broader than kaonic deuterium,
therefore, the atomic and the strong quasi-bound states cannot be misidentified.



\begin{thebibliography}{100}

\bibitem{review} N.V. Shevchenko, N.V.: Three-Body Antikaon–Nucleon Systems.
Few Body Syst. 58, 6 (2017)

\bibitem{AGS3} Alt, E.O., Grassberger, P., Sandhas, W.:
Reduction of the three-particle collision problem to multi-channel
two-particle Lippmann-Schwinger equations.
Nucl. Phys. B 2, 167 (1967)

\bibitem{Revai_Kdeu} R\'evai, J.: Three-body calculation of the $1s$ 
 level shift in kaonic deuterium with realistic $\bar{K}N$ potentials.
Phys. Rev. C 94, 054001 (2016)

\bibitem{ourKNN_I} Shevchenko, N.V., R\'evai, J.: Faddeev calculations of the $\bar{K}NN$
system with chirally-motivated $\bar{K}N$ interaction. I. Low-energy $K^- d$ scattering
and antikaonic deuterium. Phys. Rev. C 90, 034003 (2014)

\bibitem{KKN} Shevchenko, N.V., Haidenbauer, J.: Exact calculations of a quasibound
state in the $\bar{K}\bar{K}N$ system. Phys. Rev. C 92, 044001 (2015)

\bibitem{AGS4} Grassberger, P., Sandhas, W.:
Systematical treatment of the non-relativistic n-particle scattering problem.
Nucl. Phys. B 2, 181 (1967)

\bibitem{ourKNN_II} R\'evai, J., Shevchenko, N.V.: Faddeev calculations of the $\bar{K}NN$
system with chirally-motivated $\bar{K}N$ interaction. II. The $K^-pp$ quasi-bound state.
Phys. Rev. C 90, 034004 (2014)

\bibitem{PDG} Olive K.A. et al. (Particle Data Group):
The Review of Particle Physics (2015). Chin. Phys. C, 38, 090001 (2014) and 2015 update

\bibitem{Kp2exp} Sakitt, M. et al.: 
Low-energy $K^-$-meson interactions in hydrogen.
Phys. Rev. 139, B 719 (1965)

\bibitem{Kp3exp_1} Kim, J.K.:
Low-energy $K^-$-p interaction and interpretation of the 
$1405$-MeV $Y^*_0$ resonance as a $\bar{K}N$ bound state.
Phys. Rev. Lett. 14, 29 (1965)

\bibitem{Kp3exp_2} Kim, J.K.:
Multichannel phase-shift analysis of $\bar{K}N$ interaction in the region  0 to 550 MeV/c.
Phys. Rev. Lett. 19, 1074 (1967)

\bibitem{Kp4exp} Kittel, W., Otter, G., Wacek, I.: 
The $K^- $ proton charge exchange interactions at low energies and scattering lengths determination.
Phys. Lett. 21, 349 (1966)

\bibitem{Kp5exp} Ciborowski, J. et al.:
Kaon scattering and charged Sigma hyperon production in $K^- p$ interactions below 300 MeV/c.
J. Phys. G 8, 13 (1982)

\bibitem{Kp6exp} Evans, D. et al.:
Charge-exchange scattering in $K^- p$ interactions below 300 MeV/c.
J. Phys. G 9, 885 (1983)

\bibitem{gammaKp1} Tovee, D.N. et al.:
Some properties of the charged $\Sigma$ hyperons.
Nucl. Phys. B 33, 493 (1971)

\bibitem{gammaKp2} Nowak, R.J. et al.:
Charged $\Sigma$ hyperon production by $K^-$ meson interactions at rest.
Nucl. Phys. B 139, 61 (1978)

\bibitem{my_Kd} Shevchenko, N.V.: One- versus two-pole $\bar{K}N - \pi \Sigma$
potential: $K^- d$ scattering length. Phys. Rev. C 85, 034001 (2012)

\bibitem{SigmaN1} Alexander, G. et al.:
Study of the $\Lambda-N$ system in low-energy $\Lambda-p$ elastic scattering.
Phys. Rev. 173, 1452 (1968)

\bibitem{SigmaN2} Sechi-Zorn, B., Kehoe, B., Twitty, J., Burnstein, R.A.:
Low-energy $\Lambda$-proton elastic scattering.
Phys. Rev. 175, 1735 (1968)

\bibitem{SigmaN3} Eisele, F. et al.:
Elastic $\Sigma^{\pm} p$ scattering at low energies.
Phys. Lett. B 37, 204 (1971)

\bibitem{SigmaN4} Engelmann, R., Filthuth, H., Hepp, V., Kluge, E.:
Inelastic $\Sigma^-p$-interactions at low momenta.
Phys. Lett. 21, 587 (1966)

\bibitem{SigmaN5} Hepp, V., Schleich, M.: 
A new determination of the capture ratio
$r_c = \frac{\Sigma^- p \to \Sigma^0 n}{(\Sigma^- p \to \Sigma^0 n) +
 (\Sigma^- p \to \Lambda^0 n)}$, the $\Lambda^0$-lifetime
 and the $\Sigma^- - \Lambda^0$ mass difference.
Z. Phys. 214, 71 (1968)
\bibitem{VKN_SIDD12} Shevchenko, N.V.: Near-threshold $K^- d$ scattering and properties
of kaonic deuterium. Nucl. Phys. A 890-891, 50-61 (2012)

\bibitem{SIDDHARTA} Bazzi, M. {\it et al.} (SIDDHARTA Collaboration):
A new measurement of kaonic hydrogen X-rays.
Phys. Lett. B 704, 113 (2011)

\bibitem{ArgonneV18} Wiringa, R.B., Stoks, V.G.J., Schiavilla, R.:
Accurate nucleon-nucleon potential with charge-independence breaking.
Phys. Rev. C 51, 38 (1995)

\bibitem{Doles_NN} Doleschall, P.: {\it private communication}

\end{thebibliography}
\end{document}